\documentclass[a4paper,10pt]{article}
\title {Ensemble averaged entanglement of two-particle states in Fock space}
\author{
        Jan Naudts\footnote {jan.naudts@ua.ac.be}\,
  and   Tobias Verhulst\footnote{tobias.verhulst@ua.ac.be}  \\
[1ex]
\footnotesize\it
 Departement Fysica, Universiteit Antwerpen,\\
\footnotesize\it
 Groenenborgerlaan 171, 2020 Antwerpen, Belgium
} 
\date {}

\usepackage{amsfonts}
\usepackage{amsmath}

\def\Ro{{\mathbb R}}

\newcommand{\be}{\begin{eqnarray}}
\newcommand{\ee}{\end{eqnarray}}
\newcommand{\Tr}{\,{\rm Tr}\,}
\def\Io{{\mathbb I}}

\def\xx{{\bf x}}

\newtheorem{theorem}{Theorem}{}
\newtheorem{proposition}{Proposition}{}
{}
{}

\def\beginproof{\par\strut\vskip 1cm\noindent{\bf Proof}\par}
\def\endproof{\par\strut\hfill$\square$\par\vskip 0.5cm}

\begin{document}
\maketitle
\begin{abstract}
Recent results, extending the Schmidt decomposition theorem to
wavefunctions of pairs of identical particles,
are reviewed. They are used to give a definition of reduced density operators in the
case of two identical particles. Next, a method is discussed to calculate
time averaged entanglement. It is applied to a pair of identical electrons in
an otherwise empty band of the Hubbard model, and to a pair of bosons
in the Bose-Hubbard model with infinite range hopping.
The effect of degeneracy of the spectrum of the Hamiltonian
on the average entanglement is emphasised.
\end{abstract}

\section {Introduction}

\subsection* {Schmidt decomposition}

Assume that the wavefunction $\psi(\xx_1,\xx_2)$ describes
two distinguishable particles. Then there exist orthonormal bases
of wavefunctions $\phi_m(\xx_1)$ and $\chi_m(\xx_2)$ and
coefficients $p_{m}\ge 0$ such that $\psi$ can be written as
a single sum
\be
\psi=\sum_{m}\sqrt{p_m}\,\phi_m\otimes\chi_m.
\ee
This result is known as the Schmidt decomposition theorem.
See for instance \cite {NC00}, Theorem 2.7.
The reduced density matrices for each of the particles are then given by
\be
\sigma&=&\sum_mp_m |\phi_m\rangle\,\langle\phi_m|,\\
\tau&=&\sum_mp_m |\chi_m\rangle\,\langle\chi_m|.
\ee
Indeed, one verifies that for any one-particle operator $A$
\be
\Tr\sigma A
&=&\sum_mp_m\langle\phi_m|A|\phi_m\rangle\cr
&=&\sum_mp_m\langle\phi_m\otimes\chi_m|A\otimes\Io |\phi_m\otimes\chi_m\rangle\cr
&=&\langle\psi|A\otimes\Io|\psi\rangle,
\ee
and similarly
\be
\Tr\tau A&=&\langle\psi|\Io\otimes A|\psi\rangle.
\ee
The knowledge of the coefficients $p_m$ suffices to
calculate the von Neumann entropies
\be
{\cal E}(\psi)=-\Tr\sigma\ln\sigma=-\Tr\tau\ln\tau=-\sum_mp_m\ln p_m.
\label {intro:entdef}
\ee
The latter quantity is a measure for the entanglement of the two particles.

\subsection* {Identical particles}

Recently \cite {SCKLL01,GM04,GM04b}, the previous result was generalised to pairs of identical
particles, described by a wavefunction $\psi$ in a Fock space.
Let $b^\dagger(\phi)$ and $b(\phi)$ be the creation and annihilation
operators for a particle with wavefunction $\phi(\xx)$. Let $|0\rangle$
denote the vacuum state. Then for each two-particle wavefunction $\psi$ in a Fock space
there exists an orthonormal basis of wavefunctions
$\phi_m(\xx)$ in the one-particle Hilbert space
and coefficients $p_m\ge 0$
such that
\be
\psi&=&\frac 1{\sqrt 2}\sum_m\sqrt{p_m}\, b^\dagger(\phi_m)b^\dagger(\phi_m)|0\rangle,
\qquad\qquad\mbox{ (bosons)}
\label {genschmidtboson}\\
\psi&=&\sqrt 2\sum_m\sqrt{p_{2m}}\, b^\dagger(\phi_{2m})b^\dagger(\phi_{2m+1})|0\rangle,
\qquad\mbox{ (fermions)}.
\label {genschmidtfermion}
\ee
If the dimension of the one-particle Hilbert space is odd then the latter expression
does not involve all of the basis vectors $\phi_m$.

The physical interpretation of this result, in the case of bosons, is that with
probability $p_m$ the two particles are both in the same state with wavefunction $\phi_m$.
In the fermionic case, one of the particles is in the state $\phi_{2m}$, the other
in the state $\phi_{2m+1}$. It is then obvious to define reduced density matrices
$\sigma$ and $\tau$ by
\be
&&\sigma=\tau=\sum_mp_m|\phi_m\rangle\,\langle\phi_m|
\qquad\mbox{ (bosons)},
\label {bosonschmidt}\\
&&\sigma=2\sum_mp_{2m}|\phi_{2m}\rangle\,\langle\phi_{2m}|
\quad\mbox{and }\quad
\tau=2\sum_mp_{2m+1}|\phi_{2m+1}\rangle\,\langle\phi_{2m+1}|,\cr
&&\hskip 4.5cm\mbox{ (fermions)}
\label {fermionschmidt}
\ee
By convention, $p_{2n+1}=p_{2n}$ in the latter case.

In the fermion case these density matrices are far from unique
since for any pair $\phi_{2m},\phi_{2m+1}$ the
two basis vectors may be interchanged. Nevertheless, the resulting values of
the von Neumann entropies of $\sigma$ and $\tau$ are always
the same. Hence, in all cases the quantity
\be
{\cal E}(\psi)=-\sum_np_n\ln p_n
\label {entan}
\ee
can be used as a measure of entanglement.

In the next sections we reproduce the proofs of (\ref {bosonschmidt}, \ref {fermionschmidt})
and show that the eigenvalues $p_n$
of the reduced density matrices can be calculated without
actually performing the generalised Schmidt decomposition.
In this way the quantification of the entanglement of a pair of identical particles
is more easy than in the case of distinguishable particles.

\subsection* {Linear entropy}

Even the simplified method to obtain the eigenvalues $p_n$
may be too difficult for analytical treatment. For this reason we will
make use of the linear entropy instead of the von Neumann entropy (\ref {entan}).
It is still a measure of entanglement \cite {BBB06}, and is given by
\be
{\cal E}_1(\psi)=\sum_np_n(1-p_n)=1-\sum_n p_n^2.
\ee
For similar reasons the von Neumann entropy has been replaced by the
linear entropy in other papers as well, for instance in \cite {OCJQ04,LWH04,BBB06a}.

The simplification arises as follows.
Let $\rho$ be a density matrix with eigenvalues $p_n$.
Then it is often feasible to calculate $\rho^2$ by matrix multiplication
while the calculation of $\rho\ln\rho$ usually requires diagonalisation
of $\rho$. Also calculating the trace of $\rho^2$ is usually a feasible task.
The linear entropy ${\cal E}_1(\psi)$ is then obtained as $1-\Tr\rho^2$.

\subsection* {Average entanglement}

A final simplification comes from averaging the linear entanglement.
In principle, the entanglement of two particles depends on time.
Rapid fluctuations of entanglement have been reported to occur
in vibrational modes of triatomic molecules \cite {HWM06}, and between
electrons of Rydberg molecules \cite {LM06}.
They have been studied in theoretical models such as the Dicke model \cite {HH04}, a model of coupled kicked tops
\cite {DDK04}, the Harper Hamiltonian \cite {LS05}, a dimer model \cite {HCH05},
Bose-Einstein condensates \cite {XH05}.
Hence it is obvious to study the time average of the entanglement.
In \cite {NVdS06} it is shown how to replace the time average of non-linear
quantities such as the entanglement by ensemble averages.
This was applied by the present authors to study the entanglement
of distinguishable particles \cite {VN07}.

\subsection* {Overview of the paper}

The next section recalls known results about symmetric and anti-symmetric
matrices. Proofs are given in the Appendix.
The theorems of \cite {SCKLL01,GM04,GM04b} are reproduced and
the calculation of the average entanglement is explained.
Section 3 discusses the entanglement of a pair of identical electrons
in an otherwise empty band described by the Hubbard model.
Section 4 demonstrates the importance of degeneracy of the spectrum
of the Hamiltonian for a two-boson model. The paper ends with a discussion in Section 5,
followed by two Appendices.

\section {Schmidt decomposition in Fock space}

\subsection*{Known results on symmetric and anti-sym\-met\-ric matrices}

Remember that a matrix $M$ is normal if it commutes with its
hermitean conjugate $M^\dagger$. The transpose $M^T$ of
$M$ has matrix elements $(M^T)_{mn}=M_{nm}$.
The matrix $M$ is symmetric if $M^T=M$, it is anti-symmetric
if $M^T+M=0$. Any matrix with complex entries $M$ can be written as $M=V^\dagger DU$
with $D$ diagonal and with $U$ and $V$ unitary. This is the singular
decomposition of $M$. A similar result for symmetric
matrices is the following theorem. It is known as Takagi's factorisation
theorem -- see \cite {HJ88}, or \cite {GM04}, Theorem 3.4.
See \cite {EH80}, Theorem 5.5.1, for the first claim of the theorem.

\begin{theorem}
\label {thm1}
Let be given a square matrix $M$ with complex entries.
Then $M$ is symmetric if and only if it can be written as
\be
M=U^TDU
\ee
with $D$ diagonal and $U$ arbitrary.
The matrix $U$ can be chosen unitary.
\end{theorem}

Consider for example the matrix $M$, given by
\be
M=\left(\begin {array}{cc}
i &i\\
i &1        \end {array}\right).
\ee
It is not normal. Still, there exists a unitary matrix $U$,
namely
\be
U=\frac 12\left(\begin {array}{cc}
1+i &-\sqrt 2\\
1+i &\sqrt 2        \end {array}\right),
\ee
and a diagonal matrix $D=[1,1]+(1+i)[1,-1]/\sqrt 2$ such that $M=U^TDU$.
The method to find $U$ is based on the observation that
$U$ diagonalises $M^\dagger M$.
Indeed, one has
\be
M^\dagger M=U^\dagger D^\dagger DU.
\ee
This observation is essential for the calculations that follow.

The analogous result for anti-symmetric matrices is
usually formulated for matrices with real entries only.
For matrices with complex entries it follows from Lemma 1 of \cite {SCKLL01}.
As noted in \cite {LNP05}, the Theorem below is known in the Physics literature
since long --- see \cite {ZB62}.

\begin{theorem}
\label {thm2}
Let be given an anti-symmetric matrix $M$ with complex entries.
Then there exists a unitary matrix $U$ such that
$M$ can be written as $M=U^TDU$,
where $D$ has on each row and each column
at most one non-vanishing element.
\end{theorem}

If $M$ is anti-symmetric then also $D=(U^T)^\dagger MU^\dagger$ is anti-symmetric. Hence, if
$D$ has at most one non-vanishing element on each row and each column
then it can be brought into block-diagonal
form with blocks of size at most two, simply by swapping the order of rows and of columns.
This is, $D$ is a block matrix of the form
\be
D=[Z_1,Z_2, \cdots,Z_x,0,0,\cdots],
\ee
with $Z_j$ of the form
\be
Z_j=\left(\begin {array}{cc}
0 &z_1\\
-z_1 &0     \end {array}\right).
\ee

\subsection*{Application to wavefunctions in Fock space}

Take an arbitrary orthonormal basis of wavefunctions $\omega_n(\xx)$
in a finite dimensional one-particle Hilbert space.
Any two-particle wavefunction $\psi$ can be written as
\be
\psi=\sum_{mn}\lambda_{mn}\omega_m\otimes\omega_n.
\label {psiexpan}
\ee
The matrix of coefficients $\lambda_{mn}$ is denoted $\Lambda$.
In the boson case $\Lambda$ is symmetric, in the fermion case it is
anti-symmetric. Hence, by the previous theorems there exists
a unitary matrix $U$ and a matrix $D$, with at most one non-vanishing
element on each row and each column, such that $\Lambda=U^TDU$.
Then one can write
\be
\psi&=&\sum_{mnrs}U_{rm}D_{rs}U_{sn}\,\omega_m\otimes\omega_n\cr
&=&\sum_{rs}D_{rs}\phi_r\otimes\phi_s,
\label {fock:phids}
\ee
with
\be
\phi_r=\sum_mU_{rm}\omega_m.
\ee
Because the matrix $D$ has at most one non-vanishing
element on each row and each column, the double sum in (\ref {fock:phids})
reduces to a single sum. This yields
(\ref {genschmidtboson}, \ref {genschmidtfermion}, \ref {bosonschmidt}, \ref {fermionschmidt}).

Next observe that
\be
\Lambda^\dagger\Lambda=(U^TDU)^\dagger U^TDU=U^\dagger D^\dagger DU.
\ee
Hence, the matrices $\Lambda^\dagger\Lambda$ and $D^\dagger D$ have the same eigenvalues.
But the eigenvalues of $D^\dagger D$ are precisely the coefficients $p_n$
appearing in the expression (\ref {entan}) for the entanglement.
Hence, in order to calculate the entanglement of two identical systems
it suffices to expand the wavefunction $\psi$ in an arbitrary basis,
as done in (\ref {psiexpan}). Next, the matrix of expansion coefficients $\Lambda$
is used to form $\Lambda^\dagger\Lambda$. Finally, the eigenvalues $p_n$ of the latter
matrix are calculated.

\subsection* {Average entanglement using the linear entropy functional}

If now the linear entropy is used to quantify the entanglement
instead of the von Neumann entropy then one finds
\be
{\cal E}_1(\psi)=1-\Tr(\Lambda^\dagger\Lambda)^2.
\ee
Next assume that the basis of eigenvectors $\psi_n$ diagonalises
the Hamiltonian $H$. One can expand an arbitrary wavefunction $\psi$
in this basis
\be
\psi=\sum_j\sqrt{p_n}e^{i\chi_n}\psi_n,
\label {psidecomp}
\ee
with real phases $\chi_n$ and positive coefficients $p_n$ satisfying $\sum_np_n=1$.
With each basis vector $\psi_n$ corresponds an anti-symmetric matrix $\Lambda^{(n)}$
via (\ref {psiexpan}). One then obtains
\be
{\cal E}_1(\psi)=1-\sum_{mnrs}\sqrt {p_mp_np_rp_s}e^{i(\chi_n-\chi_m)}e^{i(\chi_s-\chi_r)}
\Tr\Lambda^{(m)}\strut^\dagger\Lambda^{(n)}\Lambda^{(r)}\strut^\dagger\Lambda^{(s)}.
\ee
Assume now that the spectrum of $H$ is non-degenerate.
Then the time-average entanglement of $\psi$ may be calculated as an ensemble average,
by integrating over the phase factors in the above expression. The result is
\be
\overline{{\cal E}_1(\psi)}
&=&1-\sum_{m,r}p_mp_r
\Tr\Lambda^{(m)}\strut^\dagger\Lambda^{(m)}\Lambda^{(r)}\strut^\dagger\Lambda^{(r)}\cr
& &-\sum_{m,n}p_mp_n
\Tr\Lambda^{(m)}\strut^\dagger\Lambda^{(n)}\Lambda^{(n)}\strut^\dagger\Lambda^{(m)}\cr
& &+\sum_mp_m^2\Tr\left(\Lambda^{(m)}\strut^\dagger\Lambda^{(m)}\right)^2\cr
&=&S_1(\sigma)+S_1(\tau)-\Delta,
\label {entform}
\ee
with
\be
\sigma&=&\sum_m p_m\Lambda^{(m)}\strut^\dagger\Lambda^{(m)}\\
\tau&=&\sum_n p_n\Lambda^{(n)}\Lambda^{(n)}\strut^\dagger\\
\Delta&=&1-\sum_mp_m^2\Tr\left(\Lambda^{(m)}\strut^\dagger\Lambda^{(m)}\right)^2.
\ee
Note that $\Lambda^{(m)}\strut^\dagger\Lambda^{(m)}$ and $\Lambda^{(m)}\Lambda^{(m)}\strut^\dagger$
have the same eigenvalues. Hence one has always $S_1(\sigma)=S_1(\tau)$.


The entanglement $\overline{{\cal E}_1(\psi)}$ calculated above depends on
the choice of the basis of eigenfunctions of the Hamiltonian. When the
spectrum is non-degenerate then these eigenfunctions are unique up
to a complex phase factor, which has no influence on the entanglement.
Hence the problem of non-uniqueness occurs only when the spectrum
is degenerate. In that case the decomposition (\ref {psidecomp})
of $\psi$ into eigenfunctions should be replaced by
\be
\psi=\sum_n\sqrt {p_n}\psi'_n
\ee
with
\be
\psi'_n=\frac {F_n\psi}{||F_n\psi||}
\quad\mbox{ and }\quad
p_n=|\langle\psi_n|\psi\rangle|^2.
\ee
Here, the $F_n$ are the orthogonal projections onto the degenerate eigenspaces of the
two-particle Hamiltonian.

Examples of degeneracy are discussed below.

\section {The Hubbard model}

As a first application of our method we consider the average
entanglement of a pair of identical electrons in an otherwise
empty conduction band. A suitable description is given by
the one-dimensional Hubbard model.
There is an extended literature about this model.
Its study accelerated after Lieb and Wu \cite {LW68,LW03}
showed that its spectrum can be calculated using the Bethe ansatz.
For a review paper see \cite {DEGKKK00}. In our treatment here
both electrons have the same spin. Hence, the Hamiltonian can be simplified to
\be
H=-\sum_{j,k=1}^Nt_{jk}b^\dagger_jb_k,
\label {huham}
\ee
where $b_k$ is the annihilation operator for an electron at site $k$
and the conjugate $b_k^\dagger$ is the creation operator.
The coefficients $t_{jk}$ satisfy
\be
t_{j,j+1}=t_{j,j-1}=1
\qquad \mbox{and}\qquad t_{j,k}=0\quad\mbox{ otherwise.}
\ee
Periodic boundary conditions are assumed, identifying site $N$ with site 0.

We will show that the average entanglement of the two electrons is
a non-trivial conserved quantity of this model.

\subsection*{Entanglement of the eigenvectors}

Consider a wavefunction $\psi$ describing two identical electrons, say, both with spin up,
in an otherwise empty band. Then $\psi$ is an eigenvector of $H$,
with eigenvalue $\epsilon$, 
if and only if the anti-symmetric matrix $\Lambda$ of coefficients $\lambda_{mn}$
satisfies the matrix equation
\be
T\Lambda+\Lambda T=-\epsilon\Lambda.
\ee
In the one-dimensional model with nearest neighbour interactions
(i.e., $t_{mn}=A(\delta_{m,n+1}+\delta_{m+1,n})$ and with periodic
boundary conditions (i.e., $t_{N-1,0}=t_{0,N-1}=A$) the solutions
are parameterised with two integers $r$ and $s$, with $r\not=s$,
and are given by
\be
\lambda^{(rs)}_{mn}=\frac 1{N\sqrt 2}\left[\theta(mr+ns)-\theta(nr+ms)\right]
\label {hubb1deigenv}
\ee
with $\theta(m)=\exp(2\pi im/N)$.
The corresponding eigenvalue is then
\be
E^{(rs)}=-2\Re\theta(r)-2\Re\theta(s).
\ee
Note that $\Lambda^{(rs)}=-\Lambda^{(sr)}$.

With the explicit expression (\ref {hubb1deigenv}) it is straightforward to calculate
\be
\left[\Lambda^{(rs)}\strut^\dagger\Lambda^{(rs)}\right]_{mn}
&=&\frac 1{2N^2}\sum_t\overline{\left[\theta(tr+ms)-\theta(mr+ts)\right]}\cr
& &\times
\left[\theta(tr+ns)-\theta(nr+ts)\right]\cr
&=&\frac 1{2N}\left[\overline{\theta(ms)}\theta(ns)+\overline{\theta(mr)}\theta(nr)\right].
\ee
Hence, one obtains
\be
\Tr \left[\Lambda^{(rs)}\strut^\dagger\Lambda^{(rs)}\right]^2
&=&\frac 1{4N^2}\sum_{mn}
\left[\overline{\theta(ms)}\theta(ns)+\overline{\theta(mr)}\theta(nr)\right]\cr
& &\times
\left[\overline{\theta(ns)}\theta(ms)+\overline{\theta(nr)}\theta(mr)\right]\cr
&=&\frac 12.
\ee
One concludes that all two-particle eigenvectors $\psi^{(rs)}$ are entangled,\hfill\break
with ${\cal E}_1(\psi^{(rs)})=1/2$.

One can do even more. The vectors $u^\pm$ with components
\be
u^\pm_m=\overline{\theta(mr)}\pm\overline{\theta(ms)}
\ee
are eigenvectors of the matrix $\Lambda^{(rs)}\strut^\dagger\Lambda^{(rs)}$
with eigenvalue $1/2$. All other eigenvectors have eigenvalue 0. Hence,
with the notations of previous sections the only non-vanishing eigenvalues are
$p_0=p_1=1/2$. The entanglement of the two-particle eigenvectors $\psi^{(rs)}$,
using the von Neumann entropy, is therefore
\be
{\cal E}(\psi^{(rs)})=2(-\frac 12\ln\frac 12)=\ln 2.
\ee

\subsection* {Average entanglement}

Let us now calculate the average entanglement of an arbitrary two-particle  wavefunction.
One has
\be
\Delta=1-\frac 12\sum_{rs}p^2_{rs}.
\ee
Similarly is
\be
\Tr \left[\Lambda^{(rs)}\strut^\dagger\Lambda^{(rs)}\right]\left[\Lambda^{(r's')}\strut^\dagger\Lambda^{(r's')}\right]
&=&\frac 1{4N^2}\sum_{mn}
\left[\overline{\theta(ms)}\theta(ns)+\overline{\theta(mr)}\theta(nr)\right]\cr
& &\times
\left[\overline{\theta(ns')}\theta(ms')+\overline{\theta(nr')}\theta(mr')\right]\cr
&=&\frac 14\left[\delta_{ss'}+\delta_{rr'}+\delta_{rs'}+\delta_{sr'}\right].
\ee
Hence
\be
S(\sigma)=S(\tau)=1-\frac 14\sum_{rr'ss'}p_{rs}p_{r's'}\left[\delta_{ss'}+\delta_{rr'}+\delta_{rs'}+\delta_{sr'}\right].
\ee
Using (\ref {entform}) and the normalisation condition
\be
\sum_{r>s}p_{rs}=1
\ee
one calculates
\be
\overline{{\cal E}_1(\psi)}
&=&\frac 12+\frac 12\left[\sum_{r>s}p_{rs}\right]^2+\frac 12\sum_{rs}p^2_{rs}\cr
& &-\frac 12\sum_{rr'ss'}p_{rs}p_{r's'}\left[\delta_{ss'}+\delta_{rr'}+\delta_{rs'}+\delta_{sr'}\right]\cr
&=&\frac 12+\sum_{rr'ss'}\strut'p_{rs}p_{r's'},
\ee
where the summation $\sum\strut'$ is restricted to the sets of indices $rr'ss'$
satisfying $r>r'$, $s>s'$, $r\not=s$, $r\not=s'$, $r'\not=s$, $r'\not=s'$.

In the above calculation the degeneracy of the spectrum has been neglected.
As a consequence, the result is only valid when the projection of
$\psi$ on any of the degenerate subspaces is always parallel to one of the
basis vectors $\psi^{(rs)}$. This is not the case in general.
The calculation of the entanglement of an arbitrary wavefunction is
therefore more complicated. We will not treat this general
case but end this section with an example
where degeneracy does not play.
The complications due to degeneracy will be discussed in the bosonic
example of the next section.

\subsection*{Example with $N=4$}

Take $N=4$. This means that the two electrons occupy 4 sites on a ring.
The eigenvalues are -2,0,2, each twofold degenerate.
The corresponding eigenvectors are $\psi^{(1,4)}$ and $\psi^{(3,4)}$,
$\psi^{(1,3)}$ and $\psi^{(2,4)}$, and $\psi^{(1,2)}$ and $\psi^{(2,3)}$.
We neglect the effect of the degeneracy on the average entanglement
with the argument that it can be lifted by adding a small perturbation to the model.

Let
\be
\psi=\sqrt{p}\psi^{(1,4)}+\sqrt{1-p}\psi^{(2,3)}.
\ee
Projection of $\psi$ onto the eigenspace with eigenvalue -2 gives the former term,
onto the eigenspace with eigenvalue +2 the latter term.
The average linear entanglement is
\be
\overline{{\cal E}_1(\psi)}
&=&\frac 12+p_{1,4}p_{2,3}\cr
&=&1/2+p(1-p).
\ee

\section {The bosonic model}
\label {Sboson}

As an example of the bosonic case
we consider a model which is similar to the boson-Hubbard model \cite {FWGF89,UD02,RIERS07}.

The bosonic creation and annihilation operators satisfy the commutation relations
$[b_j,b_k^\dagger]=\delta_{jk}$. The Hamiltonian is given by
\be
H=+\sum_{j,k=1}^Nt_{jk}b^\dagger_jb_k.
\label {bosonham}
\ee
However, unlike in the boson-Hubbard model, the hopping coefficients are not restricted to
nearest neighbour. They rather satisfy
\be
t_{jk}=[1-(N-1)\epsilon]\delta_{jk}+\epsilon(1-\delta_{jk}).
\ee
This model is known as the Bose-Hubbard model with infinite range hopping
\cite {BD03}.

Degeneracy is very important in this model. Indeed,
assume $\epsilon>0$. Then the ground state of the one-particle
Hamiltonian is $N-1$-fold degenerate. Hence, the two-particle
system has only three energy levels.
We will consider the state $|1,1,0,0,\cdots,0\rangle$,
in which the photons are not entangled.
Next we calculate calculate the average entanglement
and show that it tends to 1/2 when
the size $N$ of the system becomes large.

\subsection*{Projection onto invariant subspaces}

The one-particle ground state is $N-1$-fold degenerate with energy $1-N\epsilon$.
Indeed, one calculates for $m\not=n$
\be
H(b^\dagger_m-b^\dagger_n)|0\rangle
&=&\sum_j(t_{jm}-t_{jn})b^\dagger_j|0\rangle\cr
&=&(1-N\epsilon)(b^\dagger_m-b^\dagger_n)|0\rangle.
\ee
$N-1$ of these vectors $(b^\dagger_m-b^\dagger_n)|0\rangle$ are linearly independent.
The remaining eigenstate, orthogonal to the ground states, has eigenvalue 1.
Its wavefunction is
\be
\frac 1{\sqrt N}\sum_{j=1}^Nb^\dagger_j|0\rangle=b^\dagger(\phi^{(0)})|0\rangle,
\ee
with
\be
\phi^{(0)}=\frac 1{\sqrt N}\sum_j\omega_j
\ee
and $\omega_j$ the one-particle basis formed by $\omega_j=b^\dagger_j|0\rangle$.
Each of these basis vectors can be projected onto this eigenvector
\be
\omega_j=\frac 1{\sqrt N}\phi^{(0)}+\xi_j.
\ee
The vectors $\xi_j$ are orthogonal to $\phi^{(0)}$ and hence
belong to the degenerate space of eigenvectors.

The one-particle eigenfunction $\psi_0$ determines an eigenstate $\psi^{(00)}$ of the two-particle
Hamiltonian by
\be
\psi^{(00)}
=\frac 1{\sqrt 2}b^\dagger(\phi^{(0)})b^\dagger(\phi^{(0)})|0\rangle.
\ee
The initial state
\be
\psi=|1,1,0,0,\cdots\rangle=b_1^\dagger b_2^\dagger |0\rangle
\ee
is now projected onto the three invariant subspaces by writing it into the form
\be
|1,1,0,0,\cdots\rangle
&=&\frac 1Nb^\dagger(\phi^{(0)})b^\dagger(\phi^{(0)})|0\rangle\crcr
& &+\frac 1{\sqrt N}b^\dagger(\phi^{(0)})b^\dagger(\xi_1+\xi_2)|0\rangle\crcr
& &+b^\dagger(\xi_1)b^\dagger(\xi_2)|0\rangle\crcr
&\equiv&\sqrt{p^{(00)}}\psi^{(00)}
+\sqrt{p^{(11)}}\psi^{(11)}+\sqrt{p^{(01)}}\psi^{(01)},
\ee
with normalised eigenfunctions $\psi^{(\sigma,\tau)}$ and
normalisation constants $p^{(\sigma,\tau)}$.
It is straightforward to find that (see the Appendix B)
\be
p^{(00)}=\frac 2{N^2},
\qquad p^{(11)}=\left(1-\frac 1N\right)^2+\frac 1{N^2},
\qquad p^{(01)}=\frac 2N\left(1-\frac 2N\right).
\ee

\subsection*{Entanglement}

Next, one should decompose the eigenfunctions $\psi^{(00)},\psi^{(11)},\psi^{(01)}$
into the basis vectors
\be
\psi^{(\sigma,\tau)}=\sum_{jk}\lambda^{(\sigma,\tau)}_{jk}\omega_j\otimes\omega_k.
\ee
The calculation of the matrices $\Lambda^{(00)},\Lambda^{(11)},\Lambda^{(01)}$ is found in the Appendix B.
--- see (\ref {appBlam00}, \ref {appBlam11}, \ref {appBlam01}).
These are used to calculate the density matrices $\sigma$ \and $\tau$,
and the average entanglements
\be
{\cal E}(\sigma)={\cal E}(\tau)
=\frac 12+\frac 1N-\frac 2{N^2}.
\label {boson:aventsigmatau}
\ee
and
\be
\Delta
&=&\frac 12+\frac 2N-\frac {8}{N^2}+\frac {16}{N^3}-\frac {16}{N^4}.
\ee
See the Appendix B.
The final result is
\be
\overline{{\cal E}_1(\psi)}
&=&\frac 12+\frac 4{N^2}\left(1-\frac {2}{N}\right)^2.
\ee
The average entanglement is always larger than 1/2, is maximal at $N=4$ with a value of 9/16,
and converges as $1/N^2$ towards 1/2 for large $N$.

\section {Discussion}

In a rather long Introduction we have summed up a number of results
that appeared in the literature.
We have reviewed known properties of symmetric and anti-symmetric
matrices, with proofs in the Appendix A. When applied to wavefunctions
in a Fock space they lead to the definition of reduced density operators
for systems consisting of two identical particles.
These results are known. They generalise the Schmidt decomposition theorem
to pairs of identical particles. We propose to take this
generalised decomposition theorem as the basis for defining a
measure of entanglement of two identical particles.
Up to now, many authors have used for identical particles the
same expressions as for distinguishable particles.
This leads to the artificial result
that the entanglement of two identical fermions is always larger than 1.
Subtracting this constant 1 is not needed when using the definition
(\ref {intro:entdef}).

In Section 3, the technique to calculate the time-averaged entanglement
is explained. The linear entropy is used instead of the von Neumann entropy
in order to simplify the calculations. The extension of this technique
to systems of two identical particles is straightforward, using the generalised
Schmidt decomposition.

Two applications have been considered, one for fermions, the other for bosons.
In the Hubbard model the average entanglement of two identical electrons
can be calculated for arbitrary initial conditions. However, in this
calculation we have neglected the effect of degeneracy of the spectrum
of the Hamiltonian. This can be justified with the argument that
small perturbations caused by the environment would lift the degeneracy.
The average entanglement obtained in this way is always larger than one half
and is a non-trivial conserved quantity.
In the boson model the degeneracy is much worse, leaving only three distinct
energy levels. For one particular initial state we have shown that
the average entanglement can be calculated, taking degeneracy into account.
The resulting value tends to 1/2 when the size of the system becomes large.

Related results have been obtained by other authors.
L\'evay et al \cite {LNP05} consider 2 fermions in combination
with a one-particle Hilbert space of dimension 4.
Wang and Sanders \cite {WS05} use the generalised decomposition
theorem to decompose the state of the system into qubit states.
Next they calculate the entanglement
of one qubit with the others and average over the choice of qubits.
Plenio et al.\,\cite{SDP06,SDGP07,DOP07} have considered the typical
entanglement in ensembles of Gaussian states. These states differ
considerably from the two-particle states considered here.
Nevertheless, the matrix decomposition theorems might be
relevant for their context as well.

Only bipartite entanglement has been considered in the present paper.
Multipartite entanglement is more complicated and requires additional
investigation. See for instance \cite {CC06,KSTSW07,VT07}.
Neither did we study spatial entanglement of
identical particles \cite {HAV06,HBTL05,RSJ06},
or other measures of entanglement, like 
concurrence \cite {RSJ06}. Finally, note that we assume that
the time evolution is unitary. One expects that, due to interactions with the
environment, entanglement will fade away. See the review paper \cite {MCKB05}.

\section*{Appendix A}

For the sake of completeness, we give here a proof of Theorems 1 and 2.
First assume normal matrices.

\begin {proposition}
\label {normalprop}
If $M$ is normal and symmetric then there exists an orthogonal matrix $V$
and a diagonal matrix $D$ such that $M=V^TDV$.
\end {proposition}

\beginproof
Let $\{E^{(n)}\}_n$ be a spectral family in a finite dimensional Hilbert space.
Then there exists a unitary matrix $V$ and two-by-two disjunct sets $I_n$ such that
\be
E^{(n)}=V^\dagger \Io^{(n)}V
\ee
where
\be
\Io^{(n)}_{pq}&=&1\quad\mbox{ if }p=q\in I_n\cr
&=&0\quad\mbox {otherwise}.
\ee
Note that
\be
E^{(n)}_{pq}&=\sum_{r\in I_n}\overline{V_{rp}}V_{rq}.
\ee
Hence, if $E^{(n)}$ is symmetric then all elements $E^{(n)}_{pq}$ are real.
This implies that, if all $E^{(n)}$ are symmetric, then $V$ can be chosen orthogonal, i.e.~$V^\dagger=V^T$.

Let $M=\sum_n\lambda_nE^{(n)}$ be the spectral
decomposition of $M$ with all $\lambda_n$ two by two distinct.
Then also the $E^{(n)}$ are symmetric because of the uniqueness of the spectral
decomposition and because the transpose of an orthogonal projection operator
is again an orthogonal projection operator.
Hence there exists an orthogonal matrix $V$ such that
\be
M=V^TDV
\quad \mbox{ with }\quad D=\sum_n\lambda_n\Io^{(n)}.
\ee

\endproof

If $M$ is anti-symmetric then
\be
0=\sum_n\lambda_n\left[E^{(n)}+(E^{(n)})^T\right].
\ee
This does not imply that the $E^{(n)}$ are anti-symmetric (which is impossible for
a non-vanishing orthogonal projection operator anyway)!
Hence a different line of reasoning is needed.

\begin {proposition}
\label {normalprop2}
If $M$ is normal and anti-symmetric then there exists a unitary matrix $U$
such that $U^TMU$ has on each row and each column at most one non-vanishing element.
\end {proposition}

\beginproof
Let $M=\sum_n\lambda_nE^{(n)}$ be the spectral
decomposition of $M$ with all $\lambda_n$ two by two distinct.
Now assume $\zeta$ is an eigenvector of $M$ with eigenvalue $\lambda_n\not=0$,
satisfying $E^{(n)}\zeta=\zeta$.
Define $\eta$ by $\eta_r=\overline{\zeta_r}$.
Then one has
\be
(M\eta)_r
&=&\sum_sM_{rs}\eta_s\cr
&=&-\sum_sM_{sr}\overline{\zeta_s}\cr
&=&-\overline{\sum_s(M^\dagger)_{rs}\zeta_s}\cr
&=&-\overline{(M^\dagger\zeta)_r}\cr
&=&-\lambda_n\eta_r.
\ee
Hence, $\eta$ is an eigenvector of $M$ with eigenvalue $-\lambda_n$.
This implies that either $\lambda_n=0$ or there exists $m\not=n$
such that $\lambda_m=-\lambda_n$. In the latter case, $m$ and
$n$ are matching indices and $E^{(m)}$ projects on all vectors
$\eta$ obtained by taking elementwise complex conjugation
of all vectors in the range of $E^{(n)}$.

Now choose an orthonormal basis $\zeta^{(1)},\zeta^{(2)},\cdots,\zeta^{(q)}$
in the range of $E^{(n)}$ and a corresponding basis $\eta^{(1)},\eta^{(2)},\cdots,\eta^{(q)}$
in the range of $E^{(m)}$, with $\eta^{(j)}_s=\overline{\zeta^{(j)}_s}$. Do this for all
non-vanishing pairs of eigenvalues $\lambda_m=-\lambda_n$. Complement this with an
orthonormal basis in the nullspace of $M$, if present. Collect all these basis vectors
as columns of a unitary matrix $U$.
For a given $\zeta^{(j)}$ in the range of $E^{(n)}$ is,
with some abuse of notation,
\be
(U^TMU\delta_j)_p&=&(U^TMU)_{rj}\cr
&=&\sum_s(U^TM)_{rs}\zeta^{(j)}_s\cr
&=&\lambda_j\sum_s(U^T)_{rs}\zeta^{(j)}_s\cr
&=&\lambda_j\sum_s\zeta^{(r)}_s\zeta^{(j)}_s\cr
&=&\lambda_j\langle\eta^{(r)}|\zeta^{(j)}\rangle.
\ee
By construction, the latter vanishes for all but at most one value of $r$.
This ends the proof.

\endproof

Finally, the above results are generalised to arbitrary square matrices.
The argument is that found in the proof of \cite {SCKLL01}, Lemma 1.

Let be given a matrix $M$ which is either symmetric or anti-symmetric.
The matrix $MM^\dagger$ is hermitean and can be diagonalised by means of
a unitary matrix $U$, i.e.~$U^\dagger MM^\dagger U$ is diagonal.
Let $C=U^\dagger M (U^\dagger)^T$. Then $C$, like $M$, is either symmetric or
anti-symmetric. In addition it satisfies (using that $U^\dagger MM^\dagger U$ is diagonal
and that $M^T=\pm M$)
\be
C C^\dagger=U^\dagger MM^\dagger U=(U^\dagger MM^\dagger U)^T=U^TM^\dagger M(U^T)^\dagger=C^\dagger C.
\ee
This means that $C$ is normal and that, by the previous propositions,
there exists a unitary matrix $V$ such that $V^TCV$ has on each row
and each column at most one non-vanishing element.
The proof of the two theorems then follows easily.

\section*{Appendix B}

Here we present the calculation of the time
average entanglement of the initial boson state
\be
\psi=|1,1,0,\cdots,0\rangle=b_1^\dagger b_2^\dagger |0\rangle.
\ee
See Section \ref {Sboson}.

The non-degenerate eigenvector of the two-particle Hamiltonian is
\be
\psi^{(00)}
=\frac 1{\sqrt 2}b^\dagger(\phi^{(0)})b^\dagger(\phi^{(0)})|0\rangle
=\frac 1{N\sqrt 2}\sum_{j,k}b^\dagger_jb^\dagger_k|0\rangle
\ee
It has eigenvalue 2.
The projection of $|1,1,0,0,\cdots\rangle$ onto this eigenvector is
$\sqrt{p^{(00)}}\psi^{(00)}$ with $p^{(00)}= 2/N^2$.

Introduce vectors $\xi_j$, orthogonal to $\phi^{(0)}$, determined by
\be
\omega_j=\langle \phi^{(0)}|\omega_j\rangle \,\phi^{(0)}+\xi_j
=\frac 1{\sqrt N}\phi^{(0)}+\xi_j.
\ee
Then one can write
\be
|1,1,0,0,\cdots\rangle
&=&\frac 1Nb^\dagger(\phi^{(0)})b^\dagger(\phi^{(0)})|0\rangle
+b^\dagger(\xi_1)b^\dagger(\xi_2)|0\rangle\crcr
& &+\frac 1{\sqrt N}b^\dagger(\phi^{(0)})b^\dagger(\xi_1+\xi_2)|0\rangle.
\ee
The projection of $|1,1,0,0,\cdots\rangle$
onto the $(N-1)^2$-fold degenerate subspace
equals $b^\dagger(\xi_1)b^\dagger(\xi_2)|0\rangle$.
It is written as $\sqrt{p^{(11)}}\psi^{(11)}$ with
\be
p^{(11)}&=&||b^\dagger(\xi_1)b^\dagger(\xi_2)|0\rangle||^2\cr
&=&\langle\xi_1|\xi_1\rangle\,\langle\xi_2|\xi_2\rangle
+|\langle\xi_1|\xi_2\rangle|^2\crcr
&=&\left(1-\frac 1N\right)^2+\frac 1{N^2}.
\ee
The projection of $|1,1,0,0,\cdots\rangle$
onto the remaining subspace equals
\be
\frac 1{\sqrt N}b^\dagger(\phi^{(0)})b^\dagger(\xi_1+\xi_2)|0\rangle.
\ee
It is written as $\sqrt{p^{(01)}}\psi^{(01)}$ with
\be
p^{(01)}&=&
\frac 1N||b^\dagger(\phi^{(0)})b^\dagger(\xi_1+\xi_2)|0\rangle||^2\cr
&=&\frac 1N||\xi_1+\xi_2||^2\crcr
&=&\frac 2N\left(1-\frac 2N\right).
\ee

Explicit expressions for the three eigenstates are
\be
\psi^{(00)}
&=&\phi^{(0)}\otimes\phi^{(0)}\cr
&=&\frac 1N\sum_{jk}\omega_j\otimes\omega_k,\\
\psi^{(11)}&=&\frac 1{\sqrt{p^{(11)}}}\frac 1{\sqrt 2}\left(\xi_1\otimes\xi_2+\xi_2\otimes\xi_1\right)\crcr
&=&\frac 1{\sqrt{p^{(11)}}}\frac 1{\sqrt 2}\bigg[
\omega_1\otimes\omega_2+\omega_2\otimes\omega_1+\frac 2N\phi^{(0)}\otimes\phi^{(0)}\crcr
& &-\frac 1{\sqrt N}\phi^{(0)}\otimes\omega_1-\frac 1{\sqrt N}\omega_1\otimes\phi^{(0)}\crcr
& &-\frac 1{\sqrt N}\phi^{(0)}\otimes\omega_2-\frac 1{\sqrt N}\omega_2\otimes\phi^{(0)}
\bigg],\\
\psi^{(01)}&=&\frac 1{\sqrt{p^{(01)}}}\frac 1{\sqrt {2N}}\left(
\phi^{(0)}\otimes(\xi_1+\xi_2)+(\xi_1+\xi_2)\otimes \phi^{(0)}\right)\crcr
&=&\frac 1{\sqrt{p^{(01)}}}\frac 1{\sqrt {2N}}\bigg[
\phi^{(0)}\otimes(\omega_1+\omega_2)+(\omega_1+\omega_2)\otimes \phi^{(0)}\cr
& &-\frac 4{\sqrt N}\phi^{(0)}\otimes\phi^{(0)}\bigg].
\ee

The coefficients of the expansion of each of the vectors
$\psi^{(00)}$, $\psi^{(11)}$, and $\psi^{(01)}$
into the basis vectors $\omega_j\otimes\omega_k$
can be written as
\be
\lambda_{jk}^{(00)}&=&\frac 12x^{(1)}_{jk},\label {appBlam00}\\
\lambda_{jk}^{(11)}&=&\frac 1{\sqrt{p^{(11)}}}\frac 1{2N\sqrt 2}\bigg[2x^{(1)}_{jk}-2x^{(3)}_{jk}
+Nx^{(2)}_{jk}-Ny_{jk}\bigg],\label {appBlam11}\\
\lambda_{jk}^{(01)}&=&\frac 1{\sqrt{p^{(01)}}}\frac 1{N\sqrt {2}}\left[x^{(3)}_{jk}-2x^{(1)}_{jk}\right],
\label {appBlam01}
\ee
with
\be
x^{(1)}_{jk}&=&\frac 2N,\\
x^{(2)}_{jk}&=&(\delta_{j1}+\delta_{j2})(\delta_{k1}+\delta_{k2}),\\
x^{(3)}_{jk}&=&\delta_{j1}+\delta_{k1}+\delta_{j2}+\delta_{k2},\\
y_{jk}&=&(\delta_{j1}-\delta_{j2})(\delta_{k1}-\delta_{k2}).
\ee
The matrices $X^{(1)}$, $X^{(2)}$, $X^{(3)}$ span a simple Jordan algebra
of the spin factor type (see \cite {HOS84}, Section 2.9.7). The Jordan product is defined by
\be
A*B=\frac 12(AB+BA).
\ee
One verifies that
\be
X^{(1)}*X^{(1)}&=&2X^{(1)}\\
X^{(2)}*X^{(2)}&=&2X^{(2)}\\
X^{(3)}*X^{(3)}&=&2X^{(3)}+NX^{(2)}+NX^{(1)}\\
X^{(1)}*X^{(2)}&=&\frac 2NX^{(3)}\\
X^{(1)}*X^{(3)}&=&2X^{(1)}+X^{(3)}\\
X^{(2)}*X^{(3)}&=&X^{(3)}+2X^{(2)}\\
Y*Y&=&2Y\\
Y*X^{(j)}&=&0,\qquad j=1,2,3.
\ee

There exists a representation of the Jordan algebra
with the above product rules
in $\Ro^2+\Ro+\Ro$, with the product rule
\be
(u,a,\lambda)*(v,b,\mu)=(av+bu,\langle u|v\rangle+ab,\lambda\mu).
\ee
Let $u^{(1)}$ and $u^{(2)}$ be two unit vectors
satisfying $\langle u^{(1)}|u^{(2)}\rangle=-1+4/N$.
Then one can identify
\be
X^{(1)}&=&(u^{(1)},1,0)\\
X^{(2)}&=&(u^{(2)},1,0)\\
X^{(3)}&=&\frac 12(N(u^{(1)}+u^{(2)}),4,0)\\
Y&=&(0,0,2).
\ee

With this representation is
\be
\sqrt{p^{(00)}}\Lambda^{(00)}&=&\frac 1{N\sqrt 2}(u^{(1)},1,0)\\
\sqrt{p^{(11)}}\Lambda^{(11)}
&=&\frac 1{2N\sqrt 2}\left[2X^{(1)}-2X^{(3)}+NX^{(2)}-NY\right]\crcr
&=&\frac 1{2N\sqrt 2}(-(N-2)u^{(1)},N-2,-2N)\\
\sqrt{p^{(01)}}\Lambda^{(01)}
&=&\frac 1{N\sqrt 2}\left[X^{(3)}-2X^{(1)}\right]\crcr
&=&\frac 1{2N\sqrt 2}((N-4)u^{(1)}+Nu^{(2)},0,0).
\ee
It is now straightforward to calculate the squares
\be
p^{(00)}(\Lambda^{(00)})^2&=&\frac 1{N^2}(u^{(1)},1,0)\\
p^{(11)}(\Lambda^{(11)})^2&=&
\frac 1{4N^2}(-(N-2)^2u^{(1)},(N-2)^2,2N^2)\\
p^{(01)}(\Lambda^{(01)})^2&=&\frac 1{N^2}(0,N-2,0).
\ee
Summing these relations gives
\be
\sigma=\tau&=&\frac 1{4N}(-(N-4)u^{(1)},N,2N).
\ee
Squaring again gives
\be
\sigma^2=\tau^2
&=&
\frac 1{16N^2}(-2N(N-4)u^{(1)},N^2+(N-4)^2,4N^2).
\ee
The trace of the matrix represented by $(u,a,\lambda)$ equals $2a+\lambda$.
Hence, one finds (\ref {boson:aventsigmatau}).

Let us finally calculate $\Delta$.
Squaring again gives
\be
\left(p^{(00)}(\Lambda^{(00)})^2\right)^2&=&\frac 2{N^4}(u^{(1)},1,0)\\
\left(p^{(11)}(\Lambda^{(11)})^2\right)^2
&=&\frac 1{8N^4}(-(N-2)^4u^{(1)},(N-2)^4,2N^4)\\
\left(p^{(01)}(\Lambda^{(01)})^2\right)^2
&=&\frac 1{N^4}\left(0,(N-2)^2,0\right).
\ee
This gives
\be
\Delta
&=&1-\Tr\left(p^{(00)}(\Lambda^{(00)})^2\right)^2
-\Tr\left(p^{(11)}(\Lambda^{(11)})^2\right)^2
-\Tr\left(p^{(01)}(\Lambda^{(01)})^2\right)^2\crcr
&=&\frac 12+\frac 2N-\frac {8}{N^2}+\frac {16}{N^3}-\frac {16}{N^4}.
\ee

\section*{}

\end{document}